\documentclass[preprint,12pt]{elsarticle}
\usepackage[table]{xcolor}
\usepackage{lscape}
\usepackage{amsmath}
\usepackage{amssymb}
\usepackage{placeins}
\usepackage{ragged2e}
\usepackage{tabulary}
\usepackage{tabularx}
\usepackage{graphicx}
\usepackage{caption}
\usepackage{subcaption}
\usepackage{hhline}
\usepackage{longtable}
\usepackage{rotating}
\usepackage{array}
\usepackage{multicol}
\usepackage{multirow}
\usepackage{booktabs}% http://ctan.org/pkg/booktabs
\usepackage{cleveref}

\journal{arXiv}
\begin{document}
\begin{frontmatter}

\title{Profile regression for subgrouping patients with early stage Parkinson's disease}
\author{Sarini Abdullah$^1$, James McGree$^2$, Nicole White$^3$, Kerrie Mengersen$^2$, Graham Kerr$^3$}
\address{$^1$ Department of Mathematics, University of Indonesia\\ 
$^2$ARC Centre of Excellence for Mathematical and Statistical Frontiers, Queensland University of Technology (QUT), Australia \\
$^3$Institute of Health and Biomedical Innovation (IHBI), Australia\\}

\begin{abstract}
\justify
\textbf{Background:} Falls are detrimental to people with Parkinson's Disease (PD) because of the potentially severe consequences to the patients' quality of life. Thus, identifying factors that predict falls is necessary. While many studies have attempted to predict falls/non-falls, this study aimed to determine factors related to falls frequency in people with early Parkinson's disease (PD). 
\justify
\textbf{Methods:} 99 participants with early-stage PD were assessed based on two types of tests. The first type of tests is disease-specific tests, comprised of the Unified Parkinson's Disease Rating Scale (UPDRS) and the Schwab and England activities of daily living scale (SE ADL). A measure of postural instability and gait disorder (PIGD) and subtotal scores for subscales I, II, and III were derived from the UPDRS. The second type of tests is functional tests, including Tinetti gait and balance, Berg Balance Scale (BBS), Timed Up and Go (TUG), Functional Reach (FR), Freezing of Gait (FOG), Mini Mental State Examination (MMSE), and Melbourne Edge Test (MET).  Falls were recorded each month for 6 months. Clustering of patients via Finite Mixture Model (FMM) was conducted. 
\justify
\textbf{Results:}
Three clusters of patients were found: non-or single-fallers, low frequency fallers, and high frequency fallers. We identify several factors that are important to clustering PD patients. For the disease specific measures, UPDRS subscales II and III subtotals, PIGD and SE ADL are able to differentiate PD patients in non-or single-fallers from low frequency fallers. However, these factors could not differentiate PD patients with low frequency fallers from high frequency fallers. While for functional tests measures, Tinetti, TUG, and BBS turned to be important factors in clustering PD patients, and could differentiate the three clusters. 
When comparing the predictive ability of each factor to assigning new patients into cluster, factors from disease specific measures showed a stronger association with clusters than that of functional test measures. This means that disease specific measures are more informative than functional tests in explaining PD patients condition. Yet, for clustering patients with recurrent falls, functional tests complement the information from disease specific measures.

\justify
\textbf{Conclusions:}
FMM is able to cluster people with PD into three groups. We obtain several factors important to explaining the clusters and also found different role of disease specific measures and functional tests to clustering PD patients.  Upon examining these measures, it might be possible to develop new disease treatment to prevent, or to delay, the occurrence of falls.

\end{abstract}
\end{frontmatter}
\section{Introduction}

%why study falls
Falls are common in people with Parkinson's disease (PD), and can have an enormous impact on the physical and psychological health such as injury \cite{Bloem2001, Genever2005}, reduced activity levels \cite{Bloem2001} and poor quality of life \cite{Bloem2004, Romero2003}. A review of prospective studies reported that 45\% to 68\% of patients fell at least once within a 6- to 12-month period \cite{Latt2009, Paul2013,Wood2002}. Moreover, it was also reported that patients experiencing falls are more likely to fall again \cite{Pickering2007, Wood2002}. These findings highlight the importance of identifying fall risk factors to further aid clinicians design tailored treatment options to reduce falls. 

Many studies have been conducted to model falls in PD, with the attempt to discriminate fallers from non-fallers based on various measurements (see for example \cite{Catala2015, Duncan2015, Gazibara2015, Lindholm2015, Mak2010}). However, PD is progressive. Movement control becomes more debilitating over time, as the patient experiences more tremors, rigid muscles, slow movement and difficulty balancing \cite{McCoy2016}. It is plausible to hypothesize that patients might experience falls more frequently as the disease progresses. While previous studies have looked at fallers/non-fallers, this study focuses on falls frequency, with the aim to identify risk factors associated with falls frequency.

There have also been many studies aimed at classifying PD into several subtypes. Using data driven techniques, researchers have recommended several ways of subgrouping (subtypes), with the numbers of subtypes ranging from 2 to 5 \citep{Dujardin2004, Gasparoli2002, Graham1999, Lewis2005, Liu2011,Post2008, Reijnders2009,Schrag2006,Rooden2011, White2012}. The methods proposed include discriminant analysis \citep{Rooden2011,Gasparoli2002}, $K$-means clustering \citep{Liu2011}, and empirical assignment, mostly based on tremor dominance versus non-tremor dominance. Using these various methods, patients were assigned to a fixed subgroup (subtype). 

While there have been many such studies, the implementation of PD subtypes in clinical research studies has been very limited \citep{Marras2013}. It is argued that in order to be useful, the subtypes should be able to explain the disease aetiology, prognosis or treatment responsiveness \citep{Marras2013}, and should be associated with the disease progression. 

In response to this, we consider fall frequency as a measure of disease progression in addition to other clinical assessments for profiling subgroups of PD patients. Furthermore, considering the clinically highly heterogeneous characteristics of PD \citep{Lewis2005, Kehagia2010, FlensborgDamholdt2012, Marras2013}, a stochastic assignment of individuals into subgroups seems more suitable than a fixed subgroup assignment, as was done in the studies previously described. This can be addressed using finite mixture models (FMMs). Subgrouping subjects based on some covariates via the FMMs, in tandem with the associated risk factor, is known as profile regression \citep{Molitor2010, Liverani2015}.

A mixture model is an effective method of analysis in order to gain insight into patient groupings as it facilitates the identification of different sub-populations (subgroups) and their characteristics. The application of mixture models for subgrouping has a number of advantages: it quantifies the uncertainty of a patient's assignment into a given subgroup, profiles for the subgroups found can be generated using the estimated model, and it provides a statistically rigorous way of determining the number of subgroups \citep{McLachlan2004}. Examples of FMM implementation in PD studies are as demonstrated in \citep{Rolfe2012} and \citep{White2012} for the identification of PD phenotype based on symptoms. 

As our main interest is to find subgroups of patients that have similar profiles with regard to disease progression, incorporating fall frequency eases the interpretation of the subgroups produced. For example, the subgroup with a high frequency of falls is considered to be a high risk group. The characteristics of each subgroup are generated by important factors governing the subgroups and provide information on the fall risk factors. Further, upon examining these characteristics, it may be possible to develop interventions to slow the progression of PD. To the best of our knowledge, this is the first study in PD that has implemented FMMs for patient subgrouping incorporating fall frequency.

The aim of this paper is to identify risk factors related to falls frequency in people with early stages of PD using profile regression model. The focus is twofold: (i) profile generation of clusters of patients based on the disease-specific and functional tests measures which in the end, the optimal combination of those measures can be inferred to explain falls frequency and (ii) comparing the role of disease specific tests and functional tests in assigning patients into clusters. 

The remainder of this paper is organized as follows. A description of data and methodology are provided in Section \ref{methods}. Key results are  presented in Section \ref{Sec:result}, including a comparison between disease-specific measures and functional tests measures to determine the patients clustering, and generating the patients clusters profiles. A discussion of results and limitations are presented in Section \ref{Sec:discussion}. Finally, summary of overall findings is given in Section \ref{Sec:summary}.

\section{Data and Methods}
\label{methods}
\subsection{Data}
\justify

\textbf{Participants.} 101 people diagnosed with idiopathic PD participated in a prospective study conducted from March 2002 until December 2006. Participants were recruited from community support groups and neurology clinics in southeast Queensland, as a part of a larger research project conducted by the Institute of Health and Biomedical Innovation in Brisbane, Australia \cite {Kerr2010}. All participants were classified as early stage PD, determined by a Hoehn and Yahr (HY) score of 3 or less. Two patients with extremely high frequency of falls were excluded, giving a total of 99 patients data for the analysis.

\justify
\textbf{Assessments.} Each participant was followed up for a consecutive six month period with a monthly record of falls. Participants were classified as fallers if they recorded any falls during follow-up. Successful completion of falls diary was monitored by phone calls and mail correspondence. 

A series of clinical and functional tests were conducted at baseline. Participants were assessed based on two types of tests: disease specific tests and functional tests. The disease specific tests consist of the Unified Parkinson's Disease Rating Scale (UPDRS) and the Schwab and England activities of daily living scale (SE ADL). The UPDRS was assessed for three subscales: I (mentation, behaviour, mood), II (activities of daily living), and III (motor function). A measure of postural instability and gait disability (PIGD) was derived from the UPDRS (sum of items 13 - 15, 27 - 30). Sums of items in each of the UPDRS subscales yielded subtotals 1, 2, and 3, correspondingly. The functional tests consist of Tinetti (comprised of 2 subscales which relate to a clinical balance and gait), Berg Balance Scale (BBS), Timed Up and Go (TUG), Functional Reach (FR) and Freezing of Gait (FOG) for balance and gait, Mini Mental State Examination (MMSE) for cognitive impairment, and Melbourne Edge Test (MET) for visual acuity. 

\subsection{Statistical methods} \label{sec:statmethodsch5}
\subsubsection{Profile regression} 

Data on $n$ patients are denoted by $D=\{d_1, ..., d_n\}$, where $d_i$, the data of patient $i$, consists of $P$ measurements based on assessments related to the disease, $\boldsymbol{x_i}=(x_{i1},...,x_{iP})$, and fall frequency $y_i$. Each of the assessments $j$, $x_{.j}$, is assumed to follow a Gaussian distribution with mean $\mu_j$ and variance $\sigma^2_j$, that is $x_{.j} \sim N(x_j|\mu_j, \sigma_j^2)$, for $j=1, ..., P$. Fall frequency, $y_i$, is assumed to follow a Poisson distribution with mean $\theta$, $y_i \sim Po(\theta)$. Assume that patients belong to $K$ sub-populations, hereafter called subgroups. 
For patient $i$, the mixture model for the covariates is given by

\begin{equation}
f(\boldsymbol{x_i}|\pi,\boldsymbol{\mu}, \Sigma)=\sum_{k=1}^K \pi_k N_P(\boldsymbol{x_i}|\boldsymbol{\mu}_k, \Sigma_k)\,\quad i=1,...,n,
\end{equation}
where $N_P(.)$ is the $p$-dimensional Gaussian density, $\pi=\left\lbrace \pi_k\right\rbrace_{k=1}^K$ are the mixing proportions, interpreted as the probability of assigning patient $i$ with the specified criteria $\boldsymbol{x}_i$ into subgroup $k$. As proportions, each $\pi_k$ lies between 0 and 1 and $\sum_k \pi_k=1$. If patient $i$ belongs to subgroup $k$, then $\boldsymbol{\mu}_k=(\mu_{1k}, ..., \mu_{Pk})^T$ is the mean vector of $\textbf{x}_i$ with the covariance matrix of $\Sigma_k$. We assume variables are independent with a constant variance across each of the subgroups, that is, $\Sigma_k=\text{diag} (\sigma_1^2, ..., \sigma_P^2)$.

The assignment of each patient to one of the subgroups is of interest. For this purpose, a latent variable $z_i$ such that 
\begin{equation}
\boldsymbol{x_i}|z_i=k \sim N(\boldsymbol{x_i}|\mu_k, \Sigma_k), 
\end{equation}
is introduced to identify the subgroup from which each patient has been generated. This latent variable is considered as missing data, and is to be estimated as part of the model. By incorporating this latent variable, it provides an alternative interpretation of the component weights, $\pi_k=Pr(z_i=k)$, that is $\pi_k$ is the probability that object $i$ is assigned to subgroup $k$. This implies that a multinomial distribution with parameter $\pi=(\pi_1, ...,\pi_K)$ is specified for $z_i$.  

Given the response data (i.e. fall frequency), the joint covariate and response model for patient $i$ is then given by,
\begin{equation}
f(\boldsymbol{x_i},y_i|\pi,\boldsymbol{\mu}, \Sigma, \theta)=\sum_{k=1}^K \pi_k N(\boldsymbol{x_i}|\boldsymbol{\mu}_k, \Sigma_k)\,Po(y_i|\theta_k) \quad i=1,...,n,
\end{equation}
where the response $y_i$ follows a Poisson distribution with mean $\theta$ (which takes value $\theta=\theta_k$ if patient $i$ is assigned to subgroup $k$). 

The association of the profiles and the response is characterized by 
\begin{equation}
\label{eq:ch5_gamma0}
\text{log}(\theta_i|z_i)=\gamma_{z_i}+\boldsymbol{\beta}\textbf{u}_i,
\end{equation}
where $\boldsymbol{\beta}=(\beta_1,...,\beta_H)$ denotes the regression parameter coefficients associated with the covariates $\textbf{u}_i=(u_{i1}, ..., u_{iH})$, $\gamma_{z_i}$ being an individual-level intercept. In this paper, no covariates $\textbf{u}_i$ are assumed to affect fall count, and thus the model in Equation (\ref{eq:ch5_gamma0}) reduces to
\begin{equation}
\label{eq:ch5_gamma}
\text{log}(\theta_i|z_i)=\gamma_{z_i},
\end{equation}
and thus $\gamma_{z_i}$ denotes the mean fall count (in a logarithm scale) for patient $i$ if he or she belongs to subgroup $z_i$.
Hence, the likelihood for all patients is%
\begin{equation}
f(\boldsymbol{x},y|\pi, \boldsymbol{\mu}, \Sigma, \theta)=\prod_{i=1}^n{\sum_{k=1}^K \pi_k N(\boldsymbol{x_i}|\boldsymbol{\mu}_k, \Sigma_k)\, Po(y_i|\theta_k)}.
\end{equation}
If the subgroup assignment is known, then the complete data likelihood is given by,
\begin{equation}
\label{eq:ch5_complete_lik}
f(\boldsymbol{x},y|\pi, \boldsymbol{\mu}, \Sigma, \theta)=\prod_{i=1}^n \pi_{z_i} N(\boldsymbol{x_i}|\boldsymbol{\mu}_{z_i}, \Sigma_{z_i})\, Po(y_i|\theta_{z_i}).
\end{equation}

The model states that the patients, based on their symptoms represented by clinical measurements $\boldsymbol{x}$ and fall frequency $y$, are generated from $K$ distinct random processes representing the $K$ subgroups. Each of the processes is described by its own distributions, $N(\boldsymbol{x}|\boldsymbol{\mu}_k, \Sigma_k)$ and $Po(y|\theta_k)$.

We wish to make Bayesian inference for the model parameters $\boldsymbol{\Theta}$, characterized by the uncertainty in parameter values. These uncertainty are addressed though specification of prior probability distributions, as follows:
\begin{itemize}
\item $\mu_{{1,j},...,{K,j}}|\lambda_j,\Sigma_k \sim \prod_{k=1}^K {N(\mu_j, \sigma^2_j \lambda_j)}$ for means of $x_j$ in clusters $\{1,...,K\}$, $j=1, ...,p$
\item $\mu_j \sim N(0, \infty)$ for common mean of $x_j$, $j=1, ...,p$
\item $\lambda_j \sim Ga(c,d)$ for shrinkage parameter $\lambda_j$, $j=1, ...,p$
\item $\sigma_j^2 \sim IG (r,s)$ for variance of $x_j$, $j=1, ...,p$
\item $w \sim \text{Dirichlet} (\alpha_1, .., \alpha_K)$, for the mixture weight,
\end{itemize}  
where $Ga(.)$ and $IG(.)$ denote the Gamma dan Inverse-Gamma distributions respectively. 

Introduction of shrinkage parameter $\lambda_j$ is based on \cite{Yau2011}, in order to facilitate the selection of variables contributing to the clustering. Variables having the group means $\mu_{{1,j},...,{K,j}}$ shrink towards a common value $\mu_j$ are considered to be less relevant to form the clusters than other variables, and thus insight into the role of each variable could be obtained upon comparison of these parameters. 

\begin{figure}[htbp]
\centering\includegraphics[width=0.7\linewidth]{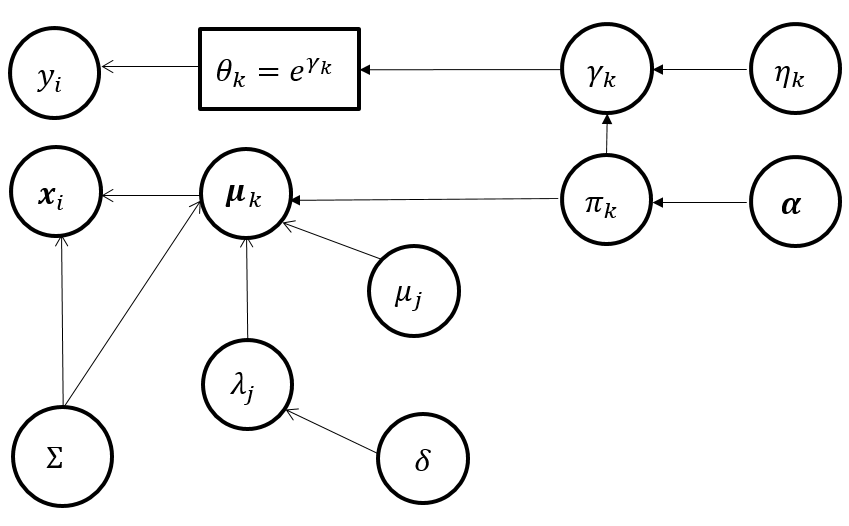}
\caption{Directed Acyclic Graph representation of the FMM model. $\Sigma=\text{diag}(\sigma_1^2, ...,\sigma_p^2)$.}
\label{fig:ch5_DAGmodel}
\end{figure}

To avoid the issue of label switching \cite{Stephens2000}, prior for the mean of falls frequency, $\theta=$ exp$(\gamma)$, for each cluster $k$ is specified as
\begin{equation*}
\gamma_1 \sim N(0,\sigma_0^2),  
\end{equation*}
and
\begin{equation*}
\gamma_k = \gamma_{k-1} + \eta_{k-1},\, k=2, ...,K
\end{equation*}
where 
$\eta_k, \, k=1,...,K-1$, are assumed to follow a truncated normal distribution at 0, i.e. $\eta_k \sim N(0,\sigma_0^2)I_{(0,\infty)}$. The indicator function $I_{(0,\infty)}$ takes the value 1 over the interval $(0,\infty)$ and 0 elsewhere. With this specification, the labeling is inline with falls frequency, where cluster with smaller label is associated with a group of patients with smaller frequency of falls. The model is visualized in Figure \ref{fig:ch5_DAGmodel}.

Having specified the likelihood and the prior distributions, using the Bayes rule, the complete data posterior is given by
\begin{align}
p(\boldsymbol{\pi}, \Phi, \theta|\textbf{x},y) & \propto \text{likelihood} \times \text{prior} \nonumber \\
   	&= f(\boldsymbol{x},y|\pi, \Phi) \times p(\Phi)p(\pi) \nonumber \\
      &= f(\textbf{x},y|\boldsymbol{\pi}, \Phi)\times p(\boldsymbol{\mu}|\lambda,\Sigma)p(\lambda)p(\Sigma)p(\theta)p(\boldsymbol{\pi}) \nonumber \\
   &=f(\textbf{x},y|\boldsymbol{\pi}, \Phi) \times \prod_{j=1}^p \left[N(\mu_j, \sigma^2_j \lambda_j)p(\lambda_j)p(\sigma_j^2)\right]p(\theta)p(\pi)   
\end{align}
where $f(\textbf{x},y|\boldsymbol{\pi}, \Phi)$ is the complete likelihood given in Equation \ref{eq:ch5_complete_lik} and the corresponding densities for the prior are as listed above. Since there is no closed form for the solution of this posterior density, samples are drawn from the posterior distribution via MCMC. 

We ran this model in WinBUGS \citep{Thomas1994}, discarding the first 10,000 burn-in iterations and sampling from the last 100,000 iterations. Exploratory data analysis prior to fitting the mixture model and post analysis were conducted in R \citep{RCoreTeam2015}. Since the number of subgroups, $K$, is assumed unknown, we fitted the models starting with $K=2$ and iteratively increased the value of $K$ until a stopping criterion was met. The  criterion is described in the following section. 

\subsection{Model diagnostics}

Questions about the number of clusters and the quality of the representation of the data often arise upon implementing a FMM. When the number of component is too large (i.e. overfitted mixture models), the overfitted latent clusters will asymptotically become empty under specific conditions for the prior of the class proportions \cite{Nasserinejad2017,Rousseau2011}. 

Various approaches have been proposed in the literature for choosing the number of latent classes in mixture models. However, no consensus has emerged regarding which of these methods performs best \cite{Nasserinejad2017}. In this study, we chose two commonly used criteria, Akaike's Information Criterion (AIC) \cite{Akaike1987}
\begin{equation*}
AIC_k=-2\text{log}f(\boldsymbol{x},y|\boldsymbol{\Theta})+2\nu_K,
\end{equation*}
and the Bayesian Information Criterion (BIC) \cite{Schwarz1978}
\begin{equation*}
AIC_k=-2\text{log}f(\boldsymbol{x},y|\boldsymbol{\Theta})+\nu_K\text{log}(n).
\end{equation*}

Deviance component, $-2\text{log}f(\boldsymbol{x},y|\boldsymbol{\Theta})$, measures the fitness of the model to data $D=\{\boldsymbol{x},y\}$ given parameter $\boldsymbol{\Theta}$. The penalty term $\nu_K$ is set to control the model's complexity by accounting for the number of parameters required to fully specify the model. Model with smaller values of $AIC_K$ and $BIC_K$ is preferred.

\subsection{Model inference}

Once the number of subgroups is selected, the aim is to make an inference on the unknown subgroup indicator, $z_i$, and subgroup parameters $\mu_k, \sigma_k^2$ representing the mean and variance of the corresponding symptoms in subgroup $k$, and $\theta_k$ representing the associated fall frequency for patients in subgroup $k$. These parameters provide information for further analysis such as profile generation for each subgroup. Moreover, assessments on the relevance and contribution of variables on the subgroup memberships will also be undertaken. 

\subsubsection{Profile generation}

Each subgroup was characterized by distinctive trends of subgroup attributes (variables). Subgroup profiles were generated by describing the trend of each variable through graphs of posterior distributions of the variable's mean in each subgroup. Furthermore, a summary of each variable , say $x_{.j}$, in each subgroup $k$ in the form of mean $\mu_{jk}$ and standard deviation $\sigma_{jk}$ will also be inspected to characterize subgroups. 
\subsubsection{Variable influence on subgroups membership}

Given the subgroup profiles, it is of interest to assess the role of each variable in assigning subjects into subgroups. 
Variable $x_{.j}$ is said to be relevant to the subgrouping if its realizations are relatively homogeneous for subjects in the same subgroup, and are different between subgroups. This relevance measurement can be inferred through the shrinkage parameter $\lambda_j$ introduced in the model in Section \ref{sec:statmethodsch5}. A high value of $\lambda_j$ indicates that $x_{.j}$ is relevant for the subgrouping, and vice versa.  Thus, relevance can be interpreted as the relative importance of variables governing the subgroups. 

More insight into the role of each variable can be obtained through the credible interval for the mean of each subgroup. The presence of non-overlapping intervals between subgroups implies there are distinct characteristics with respect to the corresponding variable. 

In addition, there will be uncertainty about this mean which shoud be considered when making comparisons. This uncertainty can be evaluated by calculating the following:

\begin{equation}
D^*_{kk'}=\frac{\sum_{t=1}^T I\{\mu_k^{(t)}>\mu_{k'}^{(t)}\}}{T}.
\end{equation}
where $T$ is the total number of MCMC iterations and $I$ is the indicator function which equals $1$ when $\mu_k>\mu_{k'}$ and 0 otherwise. $D^*_{kk'}$ measures the consistency of the distribution of $x_{.j}$ to subgroup $\{k,k'\}$. $D^*$ near 0 or 1 implies homogeneity within subgroup and a good-separation between subgroups.

To further assess a variable's influence in forming subgroups, the sensitivity of the posterior probabilities of subgroup membership with respect to changes of a given variable is examined. The presence of noticeable changes implies that the variable of interest is associated with the posterior probability of subgroup membership. Such sensitivities can also be examined via the following odds ratio

\begin{equation}
Odds_{kk'}=\frac{Pr(z_i=k|x_{ij})}{Pr(z_i=k'|x_{ij})},
\end{equation}
which states the ratio between the posterior probability of being in subgroup $k$ and the posterior probability of being in subgroup $k'$ for patient $i$ with the characteristic $x_{ij}$. Odds $>1$ suggests that patients with characteristic $x_{.j}$ are more likely to be in subgroup $k$ than in subgroup $k'$.

\section{Results}
\label{Sec:result}
\subsection{Description of participants and exploratory data}

A summary of patients' measurements and of patients classified as fallers and non-fallers is given in Table \ref{tab:Ch5desc}. Data from 99 patients were used in the analysis (66 males, 33 females, mean age 66.3 years). There was no significant association between the demographics measurements and fall/non-fall occurrences. Clear differences between fallers and non-fallers can be observed in disease specific and functional tests measurements. We will not elaborate more on these results. They are presented to show that there are differences between fallers and non-fallers, and thus motivate further elaboration on the corresponding measurements with fall frequency.

It is apparent that fall frequency increases over time, on average, as its frequency in the 6-month follow-up is higher (mean 1.4) than that in the previous year (mean 0.8). There were also patients who fell prior to participating in the study (mean 0.3), yet did not fall during the follow-up. Fall frequency for fallers in the previous year was almost 5 times higher than that of non-fallers. This indicates that once patients fell, they were prone to falling again. 
 
%\breakpage

\begin{table}[htbp]
  \centering
\caption{Summary of variables for study cohort at 6 months for all patients, and classified by fallers and non-fallers. Variables in the first part of the table (top) are for patients descriptions only (except Age). Variables for subgroup profiling are in the second part of the table (bottom). Categorical variables are summarized by their frequency (\%). Numerical variables are summarized by their means. p-value is for between groups test (Fallers and Non-fallers).}
	\scalebox{1}{
    \begin{tabular}{lcccc}
    \toprule
    \rowcolor[rgb]{ .949,  .949,  .949}      & \multicolumn{1}{l}{All patients} & Fallers & Non-fallers & p-value \\
    \midrule
    \midrule
    Fall frequency &      &      &      &  \\
\quad         In the previous year & 0.8  & 1.4  & 0.3  & $<.01$ \\
\quad        In 6 months of follow-up & 1.4  & 3    & 0    & $<.01$ \\
   % \textbf{Demographics} &      &      &      &  \\
    Gender &      &      &      & 0.5 \\
   \quad      Male & 66 (67\%)   & 38 (58\%) & 28 (42\%) &  \\
   \quad      Female & 33 (33\%)  & 15 (45 \%) & 18 (55 \%) &  \\
    Age  & 66.3 & 65.7 & 66.8 & 0.5 \\
    Height & 168.7 & 168.6 & 168.7 & 0.98 \\
    Weight & 72.8 & 74.5 & 71.4 & 0.31 \\
    Body mass index  & 25.6 & 26.2 & 25   & 0.24 \\
    MMSE & 28   & 27.8 & 28.1 & 0.53 \\
    Schwab \& England ADL & 82.2 & 79.7 & 84.5 & 0.03 \\
    Hoehn \& Yahr   &      &      &      & 0.19 \\
    \quad 1    & 27 (27\%)  & 22 (81\%) & 5 (19\%) &  \\
    \quad 2    & 44 (44\%)  & 20 (45\%) & 24 (55\%) &  \\
    \quad 3    & 28 (29\%)  & 11 (39\%) & 17 (61\%) &  \\
\hline\hline
    \textbf{Disease specifics} &      &      &      &  \\
    Duration & 6.2  & 7.3  & 5.3  & 0.05 \\
		UPDRS &&&&\\
 \quad    Subtotal 1 & 2.4  & 2.8  & 2.1  & 0.16 \\
    \quad Subtotal 2 & 10   & 11.5 & 8.8  & 0.01 \\
    \quad Subtotal 3 & 18.9 & 21.3 & 16.8 & 0.03 \\
    PIGD & 3.9  & 4.8  & 3    & $<.01$ \\
    Freezing of gait  & 4.5  & 6.1  & 3.2  & $<.01$ \\
    \textbf{Functional tests} &      &      &      &  \\
    Tinetti total & 25.9 & 24.8 & 26.8 & $<.01$ \\
    Berg balance score  & 53.6 & 52.9 & 54.2 & 0.05 \\
    Timed up and go  & 9.9  & 10.5 & 9.3  & 0.03 \\
    Functional reach (best)  & 27.4 & 27.3 & 27.5 & 0.85 \\
    \bottomrule
    \end{tabular}%
}
  \label{tab:Ch5desc}%
\end{table}%

\begin{figure}
\centering\includegraphics[width=0.9\linewidth]{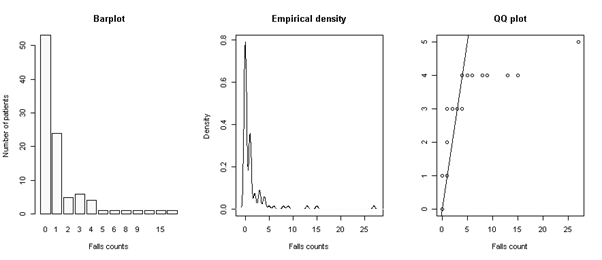}
\caption{Fall count bar plot, empirical density, and QQ-plot for all data.}
\label{fig:falls_eda1_v2}
\end{figure}

The fall counts were examined and the result is depicted in Figure \ref{fig:falls_eda1_v2}. The plot shows a very high frequency for 0/1 fall counts, a peak in the middle then low frequencies beyond a count of 5. The empirical density reveals multiple peaks. With the assumption that data follows a Poisson distribution with the mean equal to the fall count average, a QQ-plot was produced, see Figure \ref{fig:falls_eda1_v2}. The plot, however, does not support the assumption, as many points are off the line. 

\begin{figure}
\centering\includegraphics[width=0.9\linewidth]{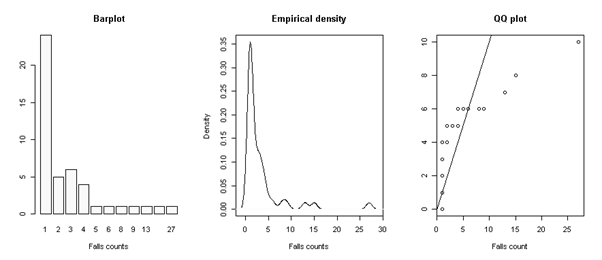}
\caption{Fall count bar plot, empirical density, and QQ-plot for data without zero fall count.}
\label{fig:falls_eda2_v2}
\end{figure}

There are excess zero counts, i.e. 52\% of the cohort did not fall. Excluding these zeros, the same process was then repeated and the result is reproduced in Figure \ref{fig:falls_eda2_v2}. The problem of multimodality still persists, and thus, the assumption of a Poisson distribution for all of the data does not appear valid. This motivates exploring whether there are underlying subgroups in the PD population, and thus motivates the consideration of profile regression models, as presented in the following section.   

\subsection{Subgrouping via profile regression models}
\label{sec:ch5_results}
\subsubsection{Model choice}

Profile regression models were fitted with the mixture component (subgroup) for the covariates ranging from 2 to 4. Based on the AIC$_K$ and BIC$_K$ criteria as presented in Table ~\ref{tab:aic}, the model with three subgroups produced the lowest values. Thus, further interpretation was based on this model. Figure \ref{fig:ch5_fallscount} presents the fall counts for subjects in the three formed subgroups, and the predicted counts (with $95\%$ credible interval) for the replicated data (Figure \ref{fig:ppc}). As indicated by the credible interval, fall counts are subgrouped to three values, very low (0 or 1), low (less than 5), and high (greater than 5), which agrees with the fall count in Figure \ref{fig:falls_eda2_v2}, implying the goodness of fit of the model.

% Table generated by Excel2LaTeX from sheet 'Sheet2'
\begin{table}[htbp]
  \centering
  \caption{Information criteria for the models with $K$ subgroups.}
	%\label{aic}
    \begin{tabular}{crrr}
    \toprule
    \multicolumn{1}{l}{Criteria} & \multicolumn{3}{c}{Number of subgroups} \\
\cmidrule{2-4}          & \multicolumn{1}{c}{2} & \multicolumn{1}{c}{3} & \multicolumn{1}{c}{4} \\
    \midrule
    \midrule
    AIC$_K$ & 7854  & 7728  & 7784 \\
    BIC$_K$ & 8007  & 7827  & 7961 \\
    \bottomrule
    \end{tabular}%
  \label{tab:aic}%
\end{table}%

\subsubsection{Subgroup description}

\begin{figure}[t!]
\centering
\begin{subfigure}[b]{.5\textwidth}
  \centering
  \includegraphics[keepaspectratio=true,scale=0.62]{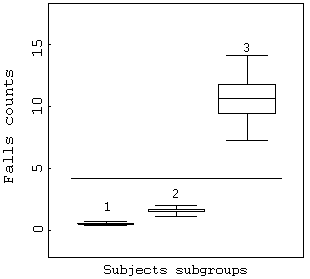}
  \caption{}
\label{fig:ch5_fallscount}
\end{subfigure}%
~
\begin{subfigure}[b]{.5\textwidth}
  \centering
  \includegraphics[keepaspectratio=true,scale=0.325]{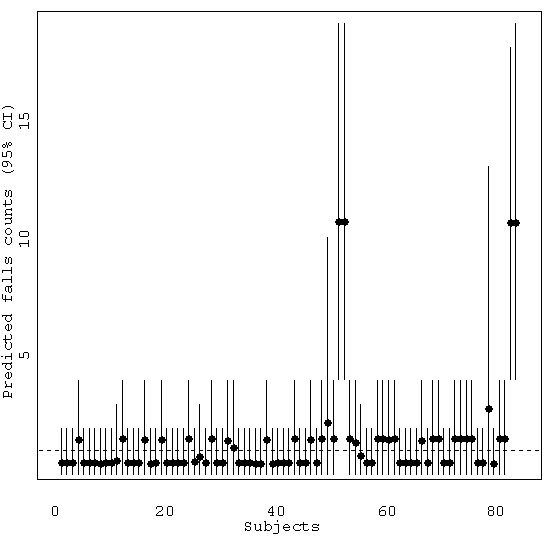}
  \caption{}
  \label{fig:ppc}
\end{subfigure}
\caption{Fall counts in three subgroups (a), Predicted fall counts with 95\% credible intervals for the replicated data (b) .}
%\label{fig2}
\end{figure}

Three subgroups were formed as shown in Figure~\ref{fig:ch5_fallscount}, with the probability of being in Subgroups 1 to 3 being 0.63, 0.27, and 0.08, respectively. Subgroup 1 consists of patients who never fall or just experienced one fall (mean fall counts = 0.5). The second subgroup is dominated by patients experiencing one or two falls (mean fall counts = 1.48). Finally, more frequent fallers were in Subgroup 3, with the mean fall counts of 10.61. 

\begin{figure}
\centering\includegraphics[width=0.5\linewidth]{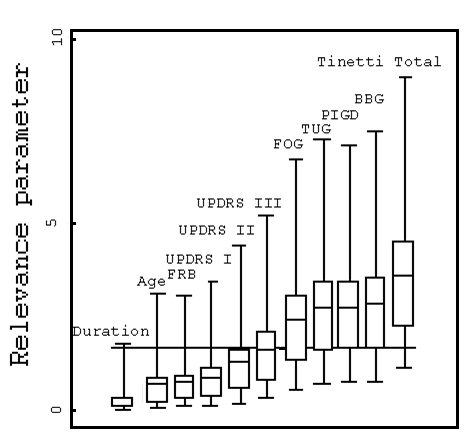}
\caption{Box plots of relative relevance of covariates used in subgrouping. Variables are ordered in ascending relevance.}
\label{fig:relevanceLambda_v3}
\end{figure}

The relevance parameter measuring the relative importance of each variable in assigning patients into the three subgroups is depicted in Figure~\ref{fig:relevanceLambda_v3}. According to this parameter, the length of time being diagnosed with the disease (Duration) was least relevant for the subgrouping, and the Tinetti total was the most relevant. Freezing of gait (FOG), timed up and go (TUG), postural instability and gait difficulty (PIGD) and balance measured by Berg balance score (BBS) shared a similar contribution to the patients' subgroupings as the relevance of these variables is about the same.

\begin{figure}[htp]
%\label{figA4}
\centering
\begin{subfigure}[b]{.3\textwidth}
  \centering
  \includegraphics[keepaspectratio=true,scale=0.35]{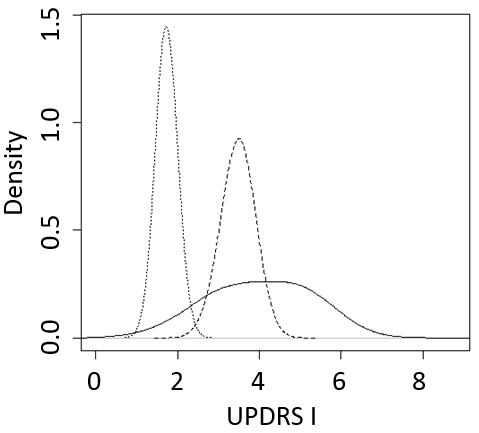}
  \caption{}
  \label{fig:updrs1clustdist}
\end{subfigure}
~
\begin{subfigure}[b]{.3\textwidth}
  \centering
  \includegraphics[keepaspectratio=true,scale=0.35]{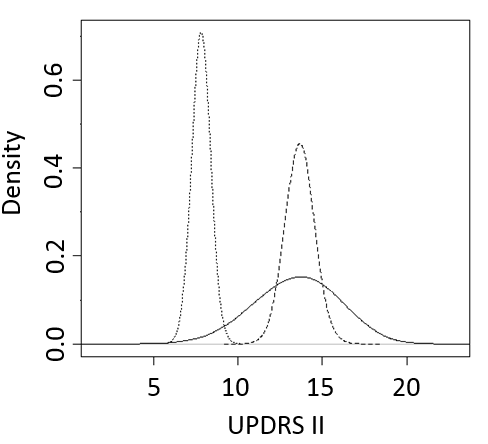}
  \caption{}
  \label{updrs2clustdist}
\end{subfigure}
~
\begin{subfigure}[b]{.3\textwidth}
  \centering
  \includegraphics[keepaspectratio=true,scale=0.35]{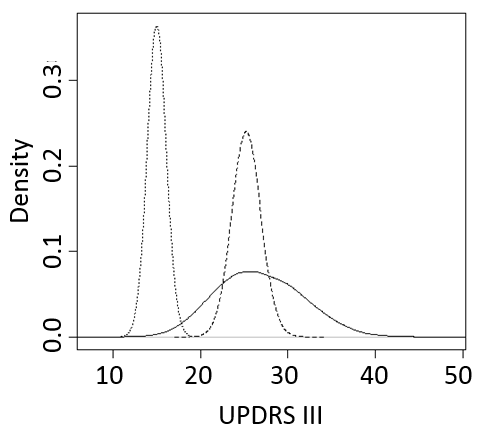}
  \caption{}
  \label{fig:updrs3clustdist}
\end{subfigure}
~
\begin{subfigure}[b]{.3\textwidth}
  \centering
  \includegraphics[keepaspectratio=true,scale=0.35]{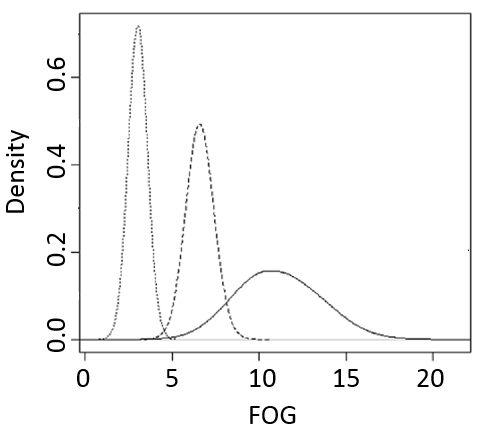}
  \caption{}
  \label{fig:fogclustdist}
\end{subfigure}%
~
\begin{subfigure}[b]{.3\textwidth}
  \centering
  \includegraphics[keepaspectratio=true,scale=0.35]{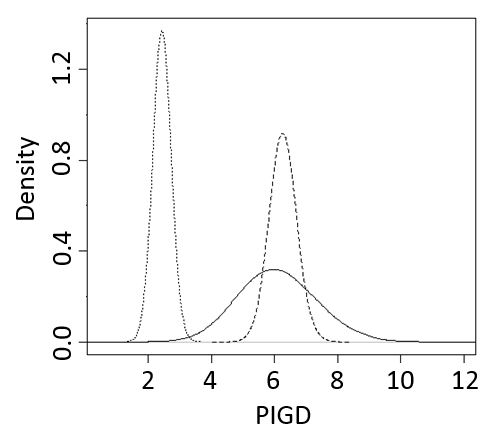}
  \caption{}
  \label{fig:pigdclustdist}
\end{subfigure}
~
\begin{subfigure}[b]{.3\textwidth}
  \centering
  \includegraphics[keepaspectratio=true,scale=0.35]{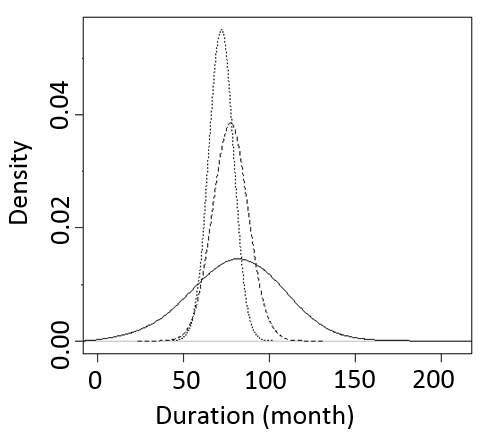}
  \caption{}
  \label{fig:durationclustdist}
\end{subfigure}
\caption{Profiles for the three subgroups -Subgroup 1 (dotted lines), Subgroup 2 (dashed lines), and Subgroup 3 (solid lines)- based on disease-specific measures.}
\label{fig:profile1}
\end{figure}

\begin{figure}[htp]
%\label{figA4}
\begin{subfigure}[b]{.3\textwidth}
  \centering
  \includegraphics[keepaspectratio=true,scale=0.32]{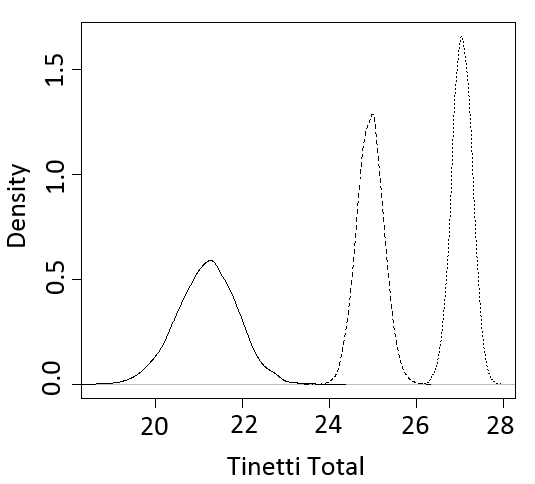}
  \caption{}
  \label{fig:Ttotalclustdist_v2}
\end{subfigure}
~
\begin{subfigure}[b]{.3\textwidth}
  \centering
  \includegraphics[keepaspectratio=true,scale=0.35]{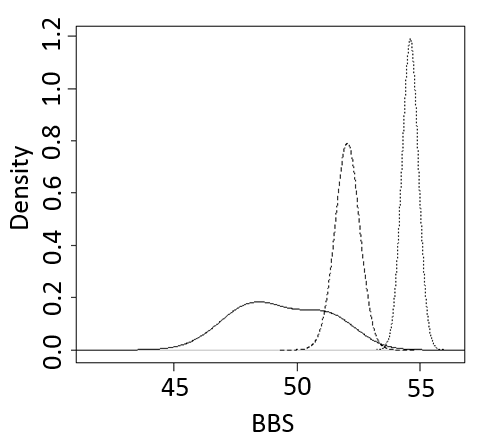}
  \caption{}
  \label{fig:bergclustdist_v2}
\end{subfigure}
~
\begin{subfigure}[b]{.3\textwidth}
  \centering
  \includegraphics[keepaspectratio=true,scale=0.35]{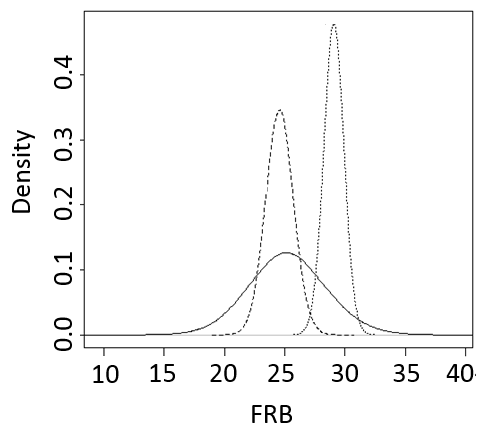}
  \caption{}
  \label{fig:frbclustdist_v2}
\end{subfigure}
~
\centering
\begin{subfigure}[b]{.3\textwidth}
  \centering
  \includegraphics[keepaspectratio=true,scale=0.35]{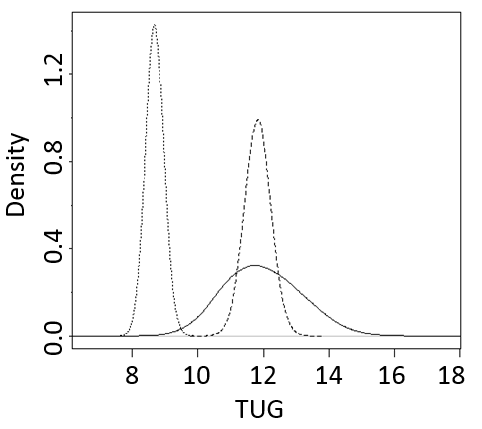}
  \caption{}
  \label{fig:tugclustdist_v2}
\end{subfigure}
~
\begin{subfigure}[b]{.3\textwidth}
  \centering
  \includegraphics[keepaspectratio=true,scale=0.35]{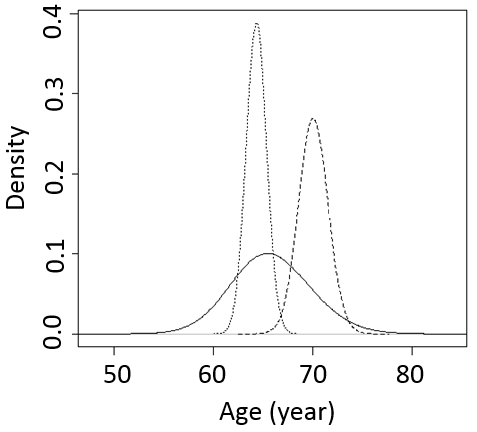}
  \caption{}
  \label{fig:ageclustdist_v2}
\end{subfigure}
\caption{Profiles for the three subgroups -Subgroup 1 (dotted lines), Subgroup 2 (dashed lines), and Subgroup 3 (solid lines)- based on functional tests and age.}
\label{fig:profile2}
\end{figure}

The posterior distribution of the means of variables was used to generate the subgroup profiles (as depicted in Figures~\ref{fig:profile1} and ~\ref{fig:profile2}). Table \ref{tab:postmeanCI} provides more insight into the contribution of each variable to the subgrouping in the form of credible intervals of posterior means and the degree of overlap of these distributions. It can be inferred that Subgroup 1 and Subgroup 2 have quite distinctive profiles based on several variables. Subgroup 3 on average is quite different from the other subgroups, however, the variation in the means is large, resulting in similar coverage values of the variables overall.

According to the disease specific measures (Figure~\ref{fig:profile1}), UPDRS subscales II, III, and PIGD can classify patients in Subgroup 1 and Subgroup 2 clearly. For UPDRS subscale I and FOG, there are slight overlaps between the two subgroups. Interestingly, the distributions are more compact in subgroups with lower fall frequency (shown by more peaked density), implying  more certainty in describing the non-or single-fallers than the frequent fallers. As in this case, it is evident from the picture that Subgroup 3, the frequent fallers, has a wide range of measures and thus cannot be differentiated with the other subgroups. On the other hand, despite there being overlap, FOG could differentiate between Subgroup 2 and Subgroup 3 more than other disease-specific measures. The trend is: high scores of the UPDRS subscales I, II, III and also of the PIGD, and worsening in freezing of gait, which indicate deterioration of the patients' condition represented by an increase in fall frequency. As far as the disease duration is concerned, it does not seem to be associated with the fall frequency. 

As for the profiles based on functional tests measures (Figure~\ref{fig:profile2}), a similar pattern with that of disease specific measures is shown: more compact distribution for posterior means of variables for Subgroup 1, clearer distinction of Subgroup 1 and Subgroup 2, and wide coverage range of variables' posterior means for patients in Subgroup 3. Among these variables, Tinetti total, TUG, and BBS best differentiated the first two subgroups. Moreover, Tinetti total can also differentiate Subgroup 3 from the other subgroups clearly, relative to the other variables. It can be inferred from the figure that better balance and gait (i.e. high scores on Tinetti and Berg balance tests) and ease in movement (lower TUG, high FRB) were associated with lower fall frequency, and vice versa.    

As indicated in Table \ref{tab:Ch5desc}, there was no significant difference in age (on average) between fallers and non-fallers. Further insight is provided by the mixture model, as depicted in Figure~\ref{fig:ageclustdist_v2}. Non-or single-fallers were relatively younger than low-frequency fallers. The non-association of age and fallers/non-fallers stated earlier was due to the contribution from recurrent fallers, whose age range covered the range for the first two subgroups. Thus, when patients were only classified as fallers or non-fallers, the age was not greatly different between the two subgroups. 

The profile for each subgroup is summarized in Table \ref{tab:postmeanCI}. The disease-specific measurements were able to differentiate Subgroup 1 and Subgroup 2 clearly, shown by non-overlapping of credible intervals of the posterior means of these variables (except for Duration). UPDRS subscales II and III, PIGD, Tinetti total, Berg balance, and timed up and go scores were completely different for the two subgroups. The degree of uncertainty measurement also provides a similar conclusion as many values of $D^*_{12}$ are either close to 0 or 1. The proportion of Subgroup 1 with the mean value of the corresponding variable is greater than that of Subgroup 2 is close to 0, or the opposite. However, these variables cannot clearly differentiate between patients with low fall frequency (Subgroup 2) and patients with high fall frequency (Subgroup 3).   

Functional test variables also show similar role to disease-specific test variables in the subgroups. These variables were very different in Subgroup 1 and Subgroup 2, and not greatly different between Subgroup 2 and Subgroup 3. 

However, further insight into $D^*_{23}$ shows that the Tinetti total, Berg balance score and freezing of gait could identify Subgroups 2 and 3 relatively well compared to the other variables. Furthermore, the Tinetti total and Berg balance score have a consistent, negative association with fall frequency $(D^*_{12}= 1, D^*_{23}\approx 1)$, which means good balance (measured by high score of the Tinetti total and BBS) is associated with less falls. As for FOG, $D^*_{23}\approx 0$ means that patients in Subgroup 2 have a lower freezing of gait score than patients in Subgroup 3. Adding the information of $D^*_{12} \approx 0$ for freezing of gait shows its consistent negative association with falls.

%\begin{landscape}
\begin{table}
\label{tab:postmeansch5}
\centering
\caption{Posterior means and 95\% credible intervals and comparison of mean differences between subgroups for each variable. Disjoint intervals are in bold. $D_{ij}^*$ is for comparing Subgroup $i$ and Subgroup $j$.}
\label{tab:postmeanCI}
\scalebox{0.8}{%
\begin{tabular}{@{}lccccc@{}}\toprule
 & Subgroup 1 & Subgroup 2 & Subgroup 3&$D_{12}^*$&$D_{23}^*$\\ \midrule
Age  & \textbf{64.3 (62.3, 66.3)} & \textbf{70.1 (67.2, 73)}& 65.8 (58.3, 74)&.0023&.7972\\
Disease-specifics\\
\quad UPDRS I & \textbf{1.7 (1.2, 2.3)} & \textbf{3.5 (2.6, 4.3)} & 4 (1.4, 6.4) &.0006 & .4695\\
\quad UPDRS II & \textbf{7.8 (6.7, 8.9)} & \textbf{13.7 (11.9, 15.4)} & 13.5 (8.4, 18.2)& 0& .584\\
\quad UPDRS III & \textbf{15 (12.9, 17.2)} & \textbf{25.3 (22.1, 28.6)} & 26.8 (17.9, 36.6)&0 & .334 \\
\quad PIGD & \textbf{2.4 (1.9, 3)} & \textbf{6.3 (5.4, 7.1)} & 6.1 (3.8, 8.6)&0 & .5052 \\
\quad Freezing of gait & \textbf{3 (2, 4.1)} & \textbf{6.6 (5.1, 8.2)} & 11.1 (6.6, 16)&.0004 & .0238 \\
\quad Duration  & 6 (4.8, 7.2) & 6.4 (4.8, 8.1) & 6.7 (2.1, 11.1)&.3211 & .4852\\
Functional tests\\
\quad Tinetti Total & \textbf{26.8 (26.1, 27.4)} &\textbf{24.6(23.7, 25.6)} & 22.7 (19.8, 25.3)& 1 & .9816 \\
\quad Berg balance score & \textbf{54.6 (54, 55.3)} & \textbf{52.1 (51.1, 53)} & 49.4 (45.9, 52.9)&1 & .9227 \\
\quad Functional reach (best) & \textbf{29.1 (27.4, 30.7)} & \textbf{24.6 (22.3, 26.8)} & 25.2 (19, 31.7)&.9974&.4278\\
\quad Timed up \& go & \textbf{8.7 (8.2, 9.2)} & \textbf{11.8 (11, 12.6)} & 12 (9.9, 14.4)&0 & .3898 \\
\bottomrule
\end{tabular}%
}
\end{table}
%\end{landscape}

\subsubsection{Uncertainty in subgroup membership}

\begin{figure}[htp]
%\label{figA4}
\centering
\begin{subfigure}[b]{.18\textwidth}
  \centering
  \includegraphics[keepaspectratio=true,scale=0.2]{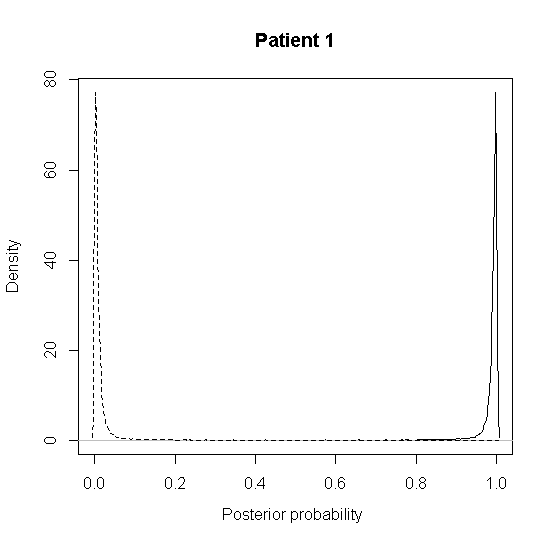}
  \caption{}
  \label{fig:pp1}
\end{subfigure}%
~
\begin{subfigure}[b]{.18\textwidth}
  \centering
  \includegraphics[keepaspectratio=true,scale=0.2]{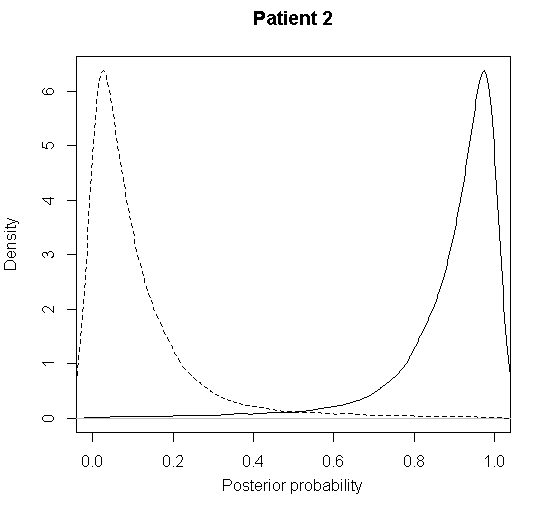}
  \caption{}
  \label{fig:pp2}
\end{subfigure}
~
\begin{subfigure}[b]{.18\textwidth}
  \centering
  \includegraphics[keepaspectratio=true,scale=0.2]{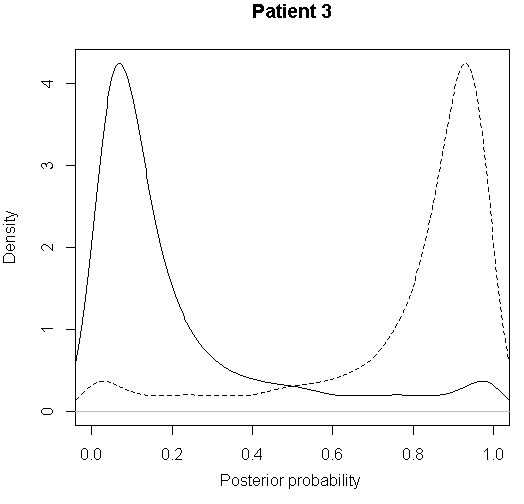}
  \caption{}
  \label{fig:pp3}
\end{subfigure}
~
\begin{subfigure}[b]{.18\textwidth}
  \centering
  \includegraphics[keepaspectratio=true,scale=0.2]{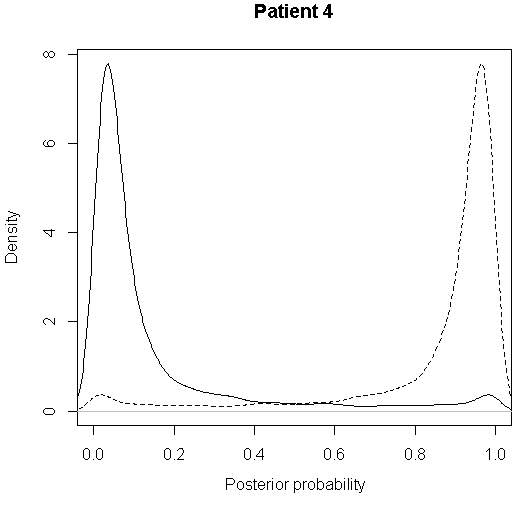}
  \caption{}
  \label{fig:pp4}
\end{subfigure}
~
\begin{subfigure}[b]{.18\textwidth}
  \centering
  \includegraphics[keepaspectratio=true,scale=0.2]{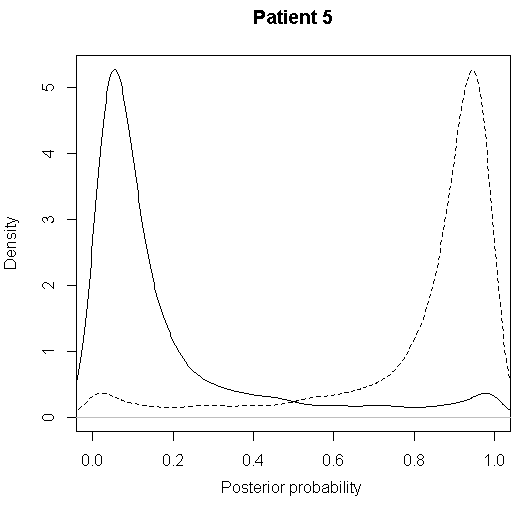}
  \caption{}
  \label{fig:pp5}
\end{subfigure}
\caption{Empirical densities of posterior probability membership for each patient described in Table~\ref{tab:modif} (solid line = subgroup 1, dashed line = subgroup 2). Subgroup 3 is not presented as the probability is almost zero for all modifications. }
\label{fig:pp}
\end{figure}

Given the subgroup profiles and relative importance of each variable, it is of interest to assess how variables determine the subgroup membership. Therefore, some variables were modified (by changing particular values, as summarized in Table ~\ref{tab:modif}), and the changes in subgroup membership were evaluated. Five modifications were implemented, and the results are shown in Figure~\ref{fig:pp}.

% Table generated by Excel2LaTeX from sheet 'Sheet1'
\begin{table}[htbp]
  \centering
  \caption{Variable modification to assess the uncertainty in subgroup membership. Patient 1 is the reference.}
\scalebox{0.8}{
    \begin{tabular}{cl}
    \toprule
    \rowcolor[rgb]{ .949,  .949,  .949} Patient & Modified variables \\
    \midrule
    \midrule
    1    & Values for age and duration diagnosed, were set to the mean for all data \\
         & Other variables were set to the corresponding means for Subgroup 1. \\
    2    & Change values for Functional test variables to their Subgroup 2 means. \\
 3  & Change values for disease-specific variables to their Subgroup 2 means.\\
 4  & Change values for Functional tests and disease-specific variables to their Subgroup 2 means.\\
 5  & Selected variables from Patient 4.\\
    \bottomrule
    \end{tabular}%
		}
  \label{tab:modif}%
\end{table}%

Considering the distinct separation of values of functional tests for Subgroup 1 and Subgroup 2 as described in the previous subsection, it is surprising that Patient 2 is assigned to Subgroup 1 (Figure~\ref{fig:pp2}). However, there is a slight change in the density of the posterior probability of membership from Patient 1 to Patient 2 (Figures~\ref{fig:pp1}, \ref{fig:pp2}), indicating a subtle effect of functional tests on subgroup membership. A more noticeable result is obtained when the disease-specific measures are changed (Patient 1 to Patient 3), as shown by the posterior probability of assigning Patient 3 to Subgroup 2 (Figures~\ref{fig:pp1}, \ref{fig:pp3}). When both functional tests and disease-specific measures are changed (Patient 4), a more certain subgroup assignment is yielded (Figure~\ref{fig:pp4}). Considering the overlap in the distribution of some measures for Subgroup 1 and Subgroup 2, omission of these variables (UPDRS I and FRB) does not change the subgroup assignment as shown for Patient 5. 

It can be inferred from Figure~\ref{fig:pp} that the contribution of disease-specific measures is stronger than that of functional tests in subgrouping -and thus in predicting fall frequency- as changes in the empirical density for Patient 1 (Figure~\ref{fig:pp1}) to that for Patient 3 (Figure~\ref{fig:pp3}) is greater than changes for Patient 2 (Figure~\ref{fig:pp2}). The effect of functional tests is noticeable when it is adjusted with the disease-specific measures (with more compact density in Figure~\ref{fig:pp4} compared to Figure~\ref{fig:pp3}). 

\begin{figure}[htp]
%\label{figA4}
\centering
\begin{subfigure}[b]{.23\textwidth}
  \centering
  \includegraphics[keepaspectratio=true,scale=0.5]{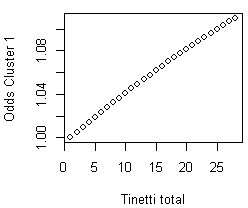}
  \caption{}
  \label{fig:pp12a}
\end{subfigure}
~
\begin{subfigure}[b]{.23\textwidth}
  \centering
  \includegraphics[keepaspectratio=true,scale=0.5]{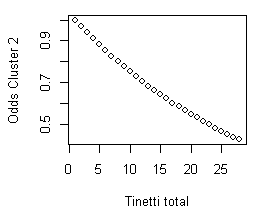}
  \caption{}
  \label{fig:pp12b}
\end{subfigure}
~
\begin{subfigure}[b]{.23\textwidth}
  \centering
  \includegraphics[keepaspectratio=true,scale=0.5]{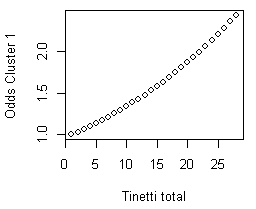}
  \caption{}
  \label{fig:pp12c}
\end{subfigure}
~
\begin{subfigure}[b]{.23\textwidth}
  \centering
  \includegraphics[keepaspectratio=true,scale=0.5]{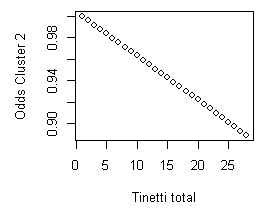}
  \caption{}
  \label{fig:pp12d}
\end{subfigure}
\caption{Odds of being in Subgroup 1 (a) and Subgroup 2 (b) when Tinetti total score is increased and all other variables were set to Subgroup 1 means. When other variables were set to Subgroup 2 means, the corresponding odds are as in (c) and (d).}
\label{fig:oddsclust12}
\end{figure}

As for the individual instrument measures, changing the value of one measure did not greatly change the posterior probability of subgroup membership (graphs not shown). However, to assess the effect of the functional tests, the odds of being in either Subgroup 1 or Subgroup 2 were calculated as the selected variable changes relative to the reference value. Balance and gait represented by the Tinetti total score was modified, taking all possible values (1 to 28), and the odds are presented in Figure~\ref{fig:oddsclust12}. 

It is shown that a patient having good balance and gait (high Tinetti total score) is more likely to be in Subgroup 1 than Subgroup 2, and vice versa. As the Tinetti total score increases, the odds of being in Subgroup 1 increases faster when other measures are set to Subgroup 2 means (Figure~\ref{fig:pp12c}) than when they are set to Subgroup 1 means (Figure~\ref{fig:pp12a}). The opposite trend occurs for the odds of being in Subgroup 2 (Figures~\ref{fig:pp12b} and \ref{fig:pp12d}). This demonstrates the association between the Tinetti and other variables used in this model.       

\begin{figure}[h]
\centering\includegraphics[width=0.9\linewidth]{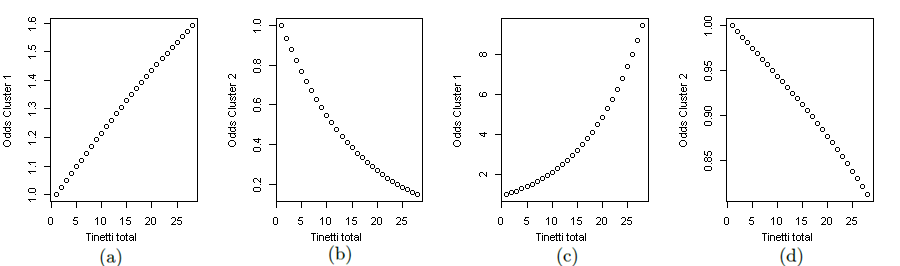}
\caption{Odds of being in Subgroup 1 (a) and Subgroup 2 (b) when the Tinetti total score is increased and TUG score is decreased, all other variables were set to Subgroup 1 means. When other variables were set to Subgroup 2 means, the corresponding odds are as in (c) and (d).}
\label{fig:oddsTinettiTUG}
\end{figure}

When changing more variables- timed up and go (TUG) scores - the changes in subgroup membership are more noticeable, as represented by the odds depicted in Figure~\ref{fig:oddsTinettiTUG}. As the Tinetti total and TUG scores reflect a better condition for the patient (higher Tinetti total, lower TUG), the odds of being in Subgroup 1 increased greatly although the other variables were set to Subgroup 2 means, representing a worse condition than Subgroup 1, (Figure \ref{fig:oddsTinettiTUG}c compared to Figure \ref{fig:oddsTinettiTUG}a). Under the same scenario, the odds of being in Subgroup 2 also diminished rapidly (Figure \ref{fig:oddsTinettiTUG}b compared to Figure \ref{fig:oddsTinettiTUG}d). 

\subsubsection{Sensitivity analysis}

\begin{figure}[htp]
%\label{figA4}
\centering
\begin{subfigure}[b]{.45\textwidth}
  \centering
  \includegraphics[keepaspectratio=true,scale=0.36]{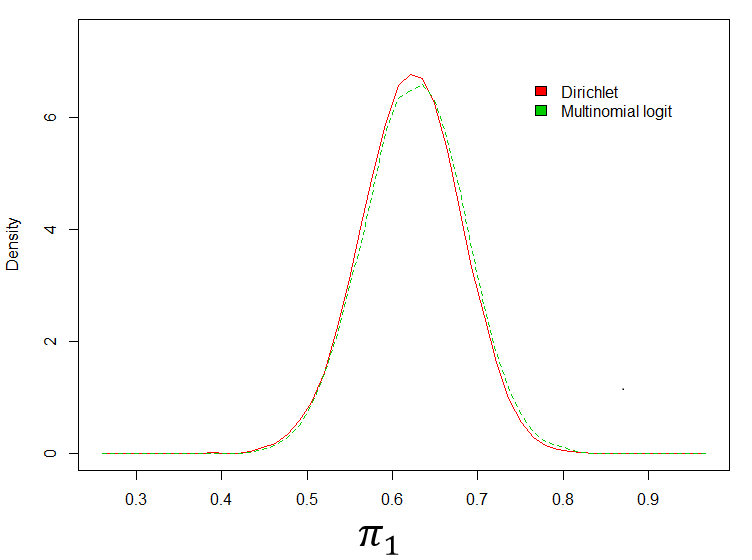}
  \caption{}
  \label{fig:ch5_sensAnalysis_piCl1}
\end{subfigure}
~
\begin{subfigure}[b]{.45\textwidth}
  \centering
  \includegraphics[keepaspectratio=true,scale=0.36]{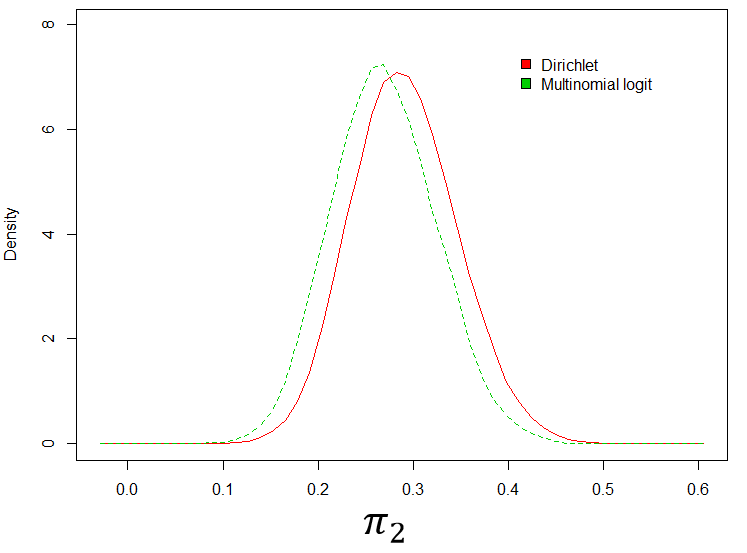}
  \caption{}
  \label{fig:ch5_sensAnalysis_piCl2}
\end{subfigure}
~

\begin{subfigure}[b]{.5\textwidth}
  \centering
  \includegraphics[keepaspectratio=true,scale=0.4]{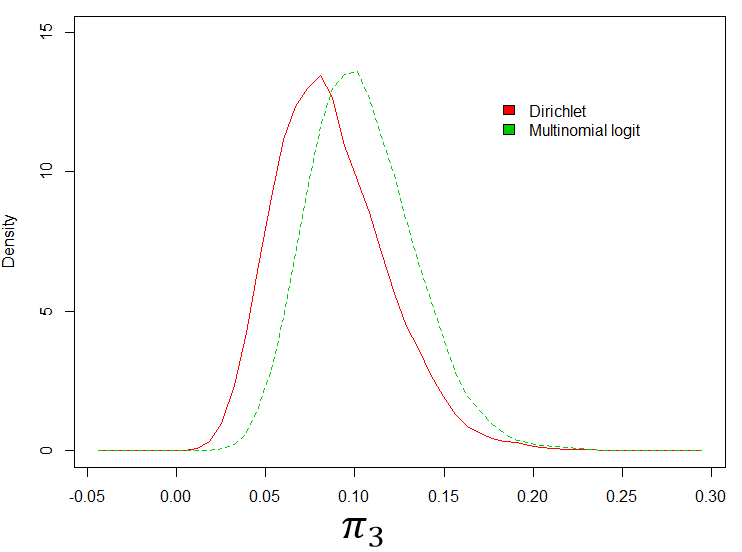}
  \caption{}
  \label{fig:ch5_sensAnalysis_piCl3}
\end{subfigure}
\caption{Posterior density of mixture weights for FMM with three subgroups:(a) Subgroup 1, (b) Subgroup 2, and (c) Subgroup 3. The red lines represent the Dirichlet prior and the green lines represent the multinomial logit prior. }
\label{fig:SensAnalysis}
\end{figure}

Two distibutions were chosen for the prior of mixture weight $\pi$: Dirichlet distribution and the multinomial logit model with a parameter which follows a Normal distribution. 
For an FMM with three classes, an uninformative Dirichlet distribution is given by $\alpha=(1,1,1)$. Denote this as Prior 1. For Prior 2, the following multinomial logit model was considered
\begin{equation}
\pi_k=\frac{\text{exp}(\gamma_k)}{\sum_k \text{exp}(\gamma_k)},
\end{equation}
with $\gamma_1=0, \gamma_k \sim N(0,1)$ for $k=2,3$. A condition of $\gamma_1 \leq \gamma_2 \leq \gamma_3$ was set to address the problem of label switching.  The posterior distribution obtained given each prior is shown in Figure \ref{fig:SensAnalysis}. 

Both priors produced similar posterior estimates for the mixture weights, indicated by the overlapping graphs of the posterior density of the mixture weights. This shows that the results are robust to the choice of prior for the mixture weight. 

\section {Discussion}
\label{Sec:discussion}

In this paper, we have demonstrated generating profiles for subgroups of patients with early stages of PD via profile regressions. Variables measuring functional and disease specific assessments were assumed to follow a Gaussian distribution, while a Poisson distribution was assumed for fall frequency. 

Three subgroups representing non- or single- fallers, low frequency fallers, and high frequency fallers were formed. Profiles characterizing each subgroup were also generated. Distinctive characteristics were identified between non-or single-fallers and low frequency fallers, while this was not the case for the high frequency fallers. For the high frequency fallers, this result needs a cautious interpretation due to the small number of cases assigned to this subgroup.

The subgroups with a higher fall frequency have wide coverage for each of the measures. This indicates that, in the early stage of PD, patients with recurrent falls cannot be differentiated from those with a low frequency of falls. Even for the duration diagnosed, there are patients who were diagnosed for a shorter time but who experienced more falls (being in Subgroup 3) than other patients. This suggests the possibility of unknown factors that need to be examined to shed light on this problem. Further, it might imply that once a patient has experienced a fall, they are prone to experience recurrent falls, regardless of other conditions represented by the above measures. 

Further, this suggests that it might be useful to identify risk factors associated with the first fall, as potentially it may be more beneficial to try to prevent that first fall rather than repeated falls. Variables that differentiate Subgroup 1 from the other subgroups offer potential to be associated with the occurrence of the first fall. The posterior density of the subgroup means reveals that the FOG, PIGD, Tinetti total and BBS are the potential risk factors for the first fall, with the Tinetti total as the strongest associated factor.  

Modifying only one variable did not change the subgroup assignment greatly. Thus, to assess the contribution of measures to subgroup membership, we modified the variables based on the disease-specific or functional tests measures. Functional test variables can differentiate well between non -or single- fallers with low frequency fallers, showing the usefulness of these measures in subgrouping the patients. However, only changing these variables had little effect on the subgroup without the change of the disease-specific measures, as has been demonstrated in Section \ref{sec:ch5_results}. This confirmed the order of importance of the variable, where functional tests are needed to enhance the information from disease-specific measures. 

Furthermore, functional tests can differentiate low frequency fallers from high frequency fallers better than disease-specific measures, as the posterior probability distribution of subgroup membership for the Tinetti total and Berg balance score had the least overlap between the two subgroups. The model with Tinetti balance and Tinetti gait as replacements to the Tinetti total was also fitted (results not shown). Tinetti balance separated Subgroup 1 and Subgroup 2 very well, but did not differentiate Subgroup 3 from the other two subgroups. Interestingly, Tinetti gait does not separate Subgroup 1 and Subgroup 2 as well as Tinetti balance (there was an overlap between the posterior probability distribution between the two subgroups). Instead, it had the least overlap between Subgroup 2 and Subgroup 3 compared to all other measures, indicating the ability of Tinetti gait to identify high frequency fallers more accurately than other variables. 

Upon examining the contribution of individual variables towards subgroup membership, the Tinetti was chosen as the measure to modify. Different rates of change on the odds of being in either Subgroup 1 or Subgroup 2 when other variables were set differently indicates the dependency of Tinetti's effect on other variables. When the patient's condition was relatively healthy (as other variable values were set to Subgroup 1 means), a worsening in balance and gait did not change the subgroup membership. However, when the condition becomes worse (changing the values to  Subgroup 2 means), the change in the Tinetti produced a noticeable effect on the subgroup membership.

This might imply that the variables change simultaneously, that is, a change in one variable would also imply a change in other variables. Thus we cannot infer the impact of one variable alone in assessing the patients' conditions. Another implication is that as the patients' conditions degenerate, slight changes of one variable could impact on their subgroup membership (which implies an increased risk of falls). 

For Tinetti, the change in subgroup membership is more explained by the balance test than the gait test in the subgroups of non-or single-fallers and low frequency fallers. The gait test gave a better explanation than the balance test when comparing low and high frequency fallers. Adding the TUG to the modification supported the results.  

An early study by \cite{Janvin2003} generated profiles for PD patients based on neuropsychological measurements, while subgroups based on the tendency towards 
delusions and hallucinations were examined in \cite{Amar2014}. A recent profiling study by \cite{Adwani2016} focused on the cognitive aspect of the disease. The results from this study provides insight into the composition of the PD population. The inclusion of fall frequency in tandem with other clinical measurements allow profiles to be generated and assessed against trends and characteristics observed in other variables. This provided additional insight into understanding the variability within a PD population.
 
\section{Summary} 
\label{Sec:summary}

Through this research study, we have identified three subgroups of patients with early stage PD, based on fall frequency, disease-specific measurements, and functional test measurements. Profiles for each subgroup were generated. Inclusion of fall frequency harnesses insight into each subgroup, namely non-or single-fallers (Subgroup 1), low frequency fallers (Subgroup 2), and high frequency fallers (Subgroup 3). Thus, a tailored treatment could be recommended based on these profiles to help prevent the deterioration of patients condition (i.e. further falls).

Using disease-specific variables, a clear differentiation of Subgroup 1 and Subgroup 2 is observed. However, these variables could not differentiate Subgroup 2 and Subgroup 3 very well. On the other hand, functional test variables were able to differentiate the 3 subgroups clearly. However, a comparison of disease specific variables and functional test variables in affecting the subgroup assignment of patients showed that the former have a higher contribution to the subgrouping than the latter. Thus, it is inferred that disease-specific measures are significant and sensitive enough to differentiate PD patients with no-or single-falls from patients with low fall frequency. Once patients have experienced at least one fall, functional tests complement the disease specific measures to signify low frequency fallers from high frequency fallers. Thus, a tailored treatment focusing on disease-specific factors could be designed to prevent the first fall, or to prevent further falls for patients who have just had one fall. For patients with recurrent falls, falls preventive treatment could be based on functional test factors. 

\bibliography{bibl_Profile_regression_for_subgrouping_PD_patients_Oct19} 

\end{document}